\documentclass[twocolumn]{aastex62}

\usepackage{graphicx}

\newcommand{\lsim}{\raisebox{-.5ex}{$\,\stackrel{\textstyle <}{\sim}\,$}}
\newcommand{\gsim}{\raisebox{-.5ex}{$\,\stackrel{\textstyle >}{\sim}\,$}}

\newcommand{\UDA}{Instituto de Astronom\'ia y Ciencias Planetarias, Universidad de Atacama, Copayapu 485, Copiap\'o, Chile}
\newcommand{\DAME}{Department of Physics and JINA Center for the Evolution of the Elements, University of Notre Dame, Notre Dame, IN 46556, USA}
\newcommand{\UNAB}{Depto. de Cs. F\'isicas, Facultad de Ciencias Exactas, Universidad Andr\'es Bello, Av. Fern\'andez Concha 700, Las Condes, Santiago, Chile}
\newcommand{\VATICAN}{Vatican Observatory, V00120 Vatican City State, Italy}
\newcommand{\UNAM}{Instituto de Astronom\'ia, Universidad Nacional Aut\'onoma de M\'exico, A.P. 70-264, 04510, Ciudad de M\'exico, Mexico}
\newcommand{\SAO}{Universidade de S\~ao Paulo, IAG, Rua do Mat\~ao 1226, Cidade Universit\'aria, S\~ao Paulo 05508-900, Brazil}     
\newcommand{\NSF}{NSF's Optical-Infrared Astronomy Research Laboratory, Tucson, AZ 85719, USA}
\newcommand{\UCN}{Instituto de Astronom\'ia, Universidad Cat\'olica del Norte, Av. Angamos 0610, Antofagasta, Chile}
\newcommand{\UDEC}{Departamento de Astronom\'\i a, Casilla 160-C, Universidad de Concepci\'on, Concepci\'on, Chile}
\newcommand{\USL}{Departamento de Astronom\'ia, Universidad de La Serena - Av. Juan Cisternas, 1200 North, La Serena, Chile}
\newcommand{\CITEVA}{Centro de Astronom\'ia (CITEVA), Universidad de Antofagasta, Avenida Angamos 601, Antofagasta 1270300, Chile}

\begin{document}

\title{Discovery of a large population of Nitrogen-Enhanced stars in the Magellanic Clouds}

\correspondingauthor{Jos\'e G. Fern\'andez-Trincado}
\email{jose.fernandez@uda.cl}

\author[0000-0003-3526-5052]{Jos\'e G. Fern\'andez-Trincado}
\affil{\UDA}

\author[0000-0003-4573-6233]{Timothy C. Beers}
\affil{\DAME}

\author[0000-0002-7064-099X]{Dante Minniti}
\affil{\UNAB}
\affil{\VATICAN}

\author{Leticia Carigi}
\affil{\UNAM}

\author[0000-0001-9264-4417]{Beatriz Barbuy}
\affil{\SAO}	

\author[0000-0003-4479-1265]{Vinicius M. Placco}	
\affil{\DAME}
\affil{\NSF}

\author{Christian Moni Bidin}
\affil{\UCN}

\author{Sandro Villanova}
\affil{\UDEC}	

\author[0000-0002-1379-4204]{Alexandre Roman-Lopes}
\affil{\USL}

\author[0000-0003-4752-4365]{Christian Nitschelm}
\affil{\CITEVA}

\begin{abstract}
We report the APOGEE-2S$+$ discovery of a unique collection of nitrogen-enhanced mildly metal-poor giant stars, peaking at [Fe/H]$\sim -0.89$ with no carbon enrichment, toward the Small and Large Magellanic Clouds (MCs), with abundances of light- (C, N), odd-Z (Al, K) and $\alpha-$elements (O, Mg, Si) that are typically found in Galactic globular clusters (GCs). Here we present 44 stars in the MCs that exhibit significantly enhanced [N/Fe] abundance ratios, well above ([N/Fe]$\gsim +0.6$) typical Galactic levels at similar metallicity, and a star that is very nitrogen-enhanced ([N/Fe]$> +2.45$). Our sample consists of luminous evolved stars on the asymptotic giant branch (AGB), eight of which are classified as bonafide semi-regular (SR) variables, as well as low-luminosity stars similar to that of stars on the tip of the red giant branch of stellar clusters in the MCs. It seems likely that whatever nucleosynthetic process is responsible for these anomalous MC stars it is similar to that which caused the common stellar populations in GCs. We interpret these distinctive C-N patterns as the observational evidence of the result of tidally shredded GCs in the MCs. These findings might explain some previous conflicting results over bulge N-rich stars, and broadly help to understand GC formation and evolution. Furthermore, the discovery of such a large population of N-rich AGB stars in the MCs suggests that multiple stellar populations might not only be exotic events from the past, but can also form at lower redshift.
\end{abstract}

\section{Introduction} 
\label{section1}

It has been well-established that some metal-poor ($ -2.0<$[Fe/H]$< -0.7$) stars within the Milky Way may produce a [N/Fe] over-abundance \citep[i.e., N-rich stars;][]{Bessell1982, Beveridge1994, Johnson2007, Fernandez-Trincado2016, Martell2016, Fernandez-Trincado2017, Schiavon2017, Fernandez-Trincado2019, Fernandez-Trincado2019a, Fernandez-Trincado2019b, Fernandez-Trincado2020} that replicates or exceeds the chemical patterns seen in the so-called \textit{second-generation}\footnote{\textit{Second-generation} is used here to refer to groups of stars in the halo and star clusters that display altered (i.e., different to those of MW field stars) light- element abundances (He, C, N, O, Na, Al, and Mg).} stars in GCs in the bulge and halo of the Milky Way (MW) \citep[][]{Fernandez-Trincado2019c, Szabolcs2020}, and as such we would expect to find similar chemical signatures in the Large Magellanic Cloud (LMC), that also has a halo \citep{Minniti2003, Borissova2006}. On the contrary, for metal-poor field stars with metallicities down to [Fe/H]$=-0.7$, the [N/Fe] ratio is not higher than $+$0.5. 

Galactic GCs are expected to lose mass through processes like evaporation and tidal stripping \citep[e.g.,][]{Baumgardt2003}, and the favored hypothesis for the origin of the N-rich stars is that they were once part of a GC. This conjecture appears to be well-supported by the chemo-kinematics similarity between a number of nitrogen-enriched MW field stars and its possible GC progenitors \citep[see, e.g.,][]{Martell2016, Tang2020, Fernandez-Trincado2019, Fernandez-Trincado2019, Fernandez-Trincado2020}. Alternatively, it has been suggested that such a N-rich population could include the oldest stars in the MW, which could have been born in high-density environments \citep{Chiappini2011, Bekki2019}.

The N-rich population is often typified by larger N over-abundances, accompanied by decreased abundances of carbon ([C/Fe]$\lesssim+0.15$) and $\alpha-$elements (O, and Mg). Sometimes they exhibit atmospheres extremely enriched in aluminum ([Al/Fe]$>+0.5$) and \textit{s}-process elements, suggesting that some of them could be objects in the AGB evolutionary stages that have undergone the hot bottom burning \citep[see e.g.,][]{Fernandez-Trincado2016, Pereira2017, Fernandez-Trincado2017, Fernandez-Trincado2019}, including a few cases which became strongly enriched in phosphorus \citep[see, e.g.,][]{Masseron2020}, which could be biased towards red giant branch (RGB) stars in previous studies \citep{Schiavon2017}, or could be objects chemically enriched by an AGB companion \citep[e.g.,][]{Cordero2015, Fernandez-Trincado2019b}.

The existence of N-rich stars in specific environments has proven to have important implications in the chemical makeup of multiple populations (MPs) in the context of GC evolution across a wide range of metallicity \citep[see, e.g.,][]{Renzini2015, Fernandez-Trincado2019c, Szabolcs2020}. Establishing whether N-rich stars in the MW and/or nearby Local Group systems formed in a GC could either reveal that MP populations can also form, to some extent, in lower-density environments \citep[e.g.,][]{Savino2019}, as well as provide important insights on the assembly history of their host systems, in particular on the role of GC disruption.

While the abundance properties of N-rich stars have been limited to GC stars and the inner/outer (up to $\lesssim 100$ kpc) stellar halo \citep{Martell2016, Fernandez-Trincado2017, Fernandez-Trincado2019, Fernandez-Trincado2019a} and widely explored, little is known of these stars in nearby dwarf galaxies that surround the MW, even though some attempts to search for dissolved GCs in these systems via the investigation of chemical anomalies have been tried \citep[e.g.,][]{Lardo2016}. 

The Apache Point Observatory Galactic Evolution Environment \citep[APOGEE:][]{Majewski2017}, is currently obtaining near-IR spectra for stars in the Small and Large Magellanic (SMC/LMC) Clouds \citep{Nidever2020}, providing us with an excellent window to examine the presence of disrupted GCs in nearby Local Group galaxies. In this Letter, we report the discovery of a large population of N-rich stars likely associated with GC dissolution and/or evaporation in the SMC/LMC. To our knowledge, none of the large spectroscopic surveys of the SMC/LMC system have so far included measurements of nitrogen abundances. 

This work is organized as follows. In Section \ref{section2}, we discuss the data and selection criteria employed to create the parent stellar sample used throughout the paper. We present our analysis and conclusions in Section \ref{section3}.  

\begin{figure*}[t]
	\begin{center}
		\includegraphics[height = 19. cm]{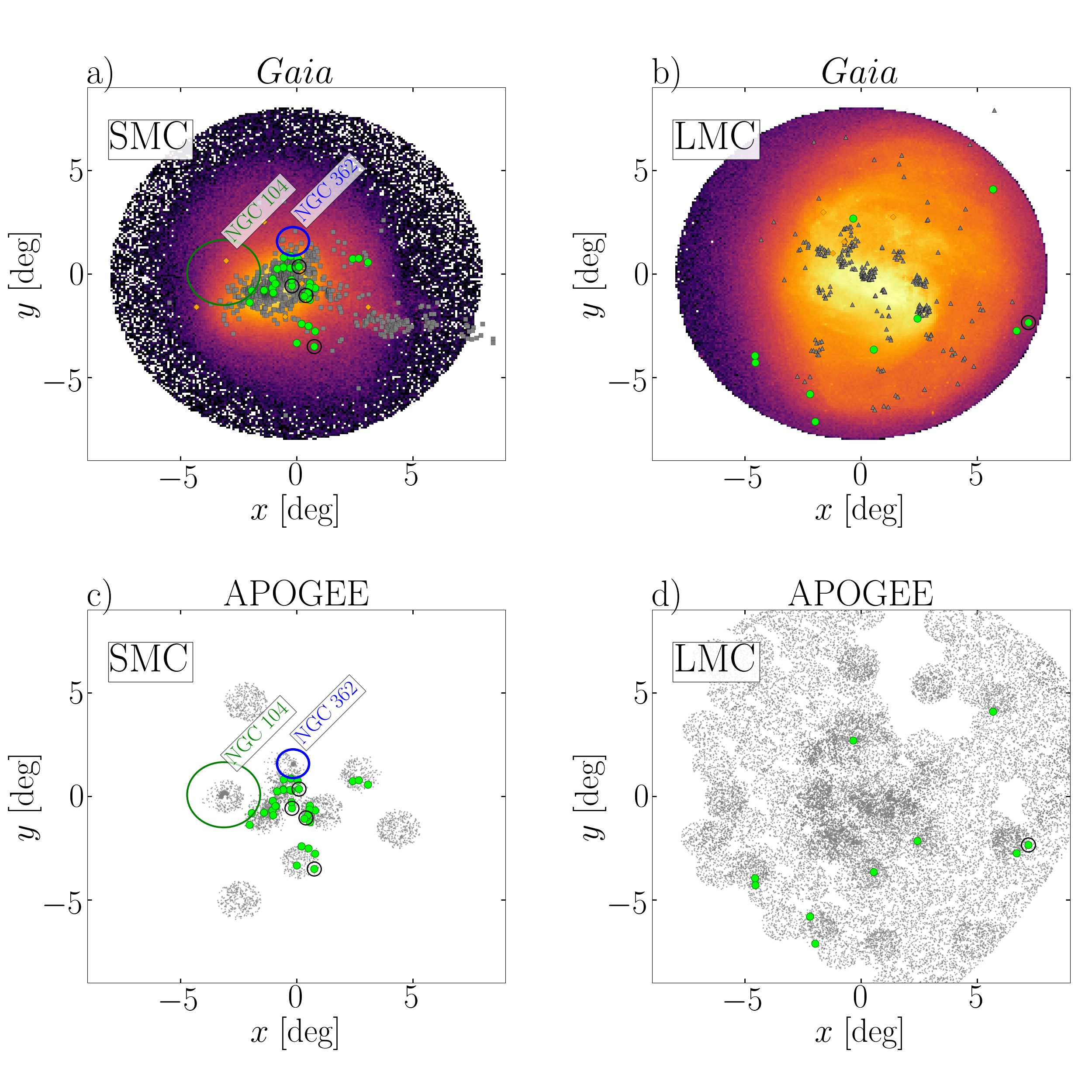}\\		
		\caption{Density distribution on the sky of the stars selected as members of the SMC (panel a) and LMC (panel b) from the \citet{Helmi2018}, and the stars in the APOGEE 2S$+$ footprint are shown in panels (c) and (d). The N-rich stars are marked as lime dot symbols. MC clusters from \citet{Bica1995}, \citet{Palma2016}, and \citet{Milone2020} are marked as gray squares, gray triangles, and orange diamond symbols, respectively. The large blue and green circles mark the tidal radius of NGC 362 and NGC 104, respectively.}
		\label{fig1}
	\end{center}
\end{figure*}

\begin{figure*}[t]
	\begin{center}
		\includegraphics[height = 13. cm]{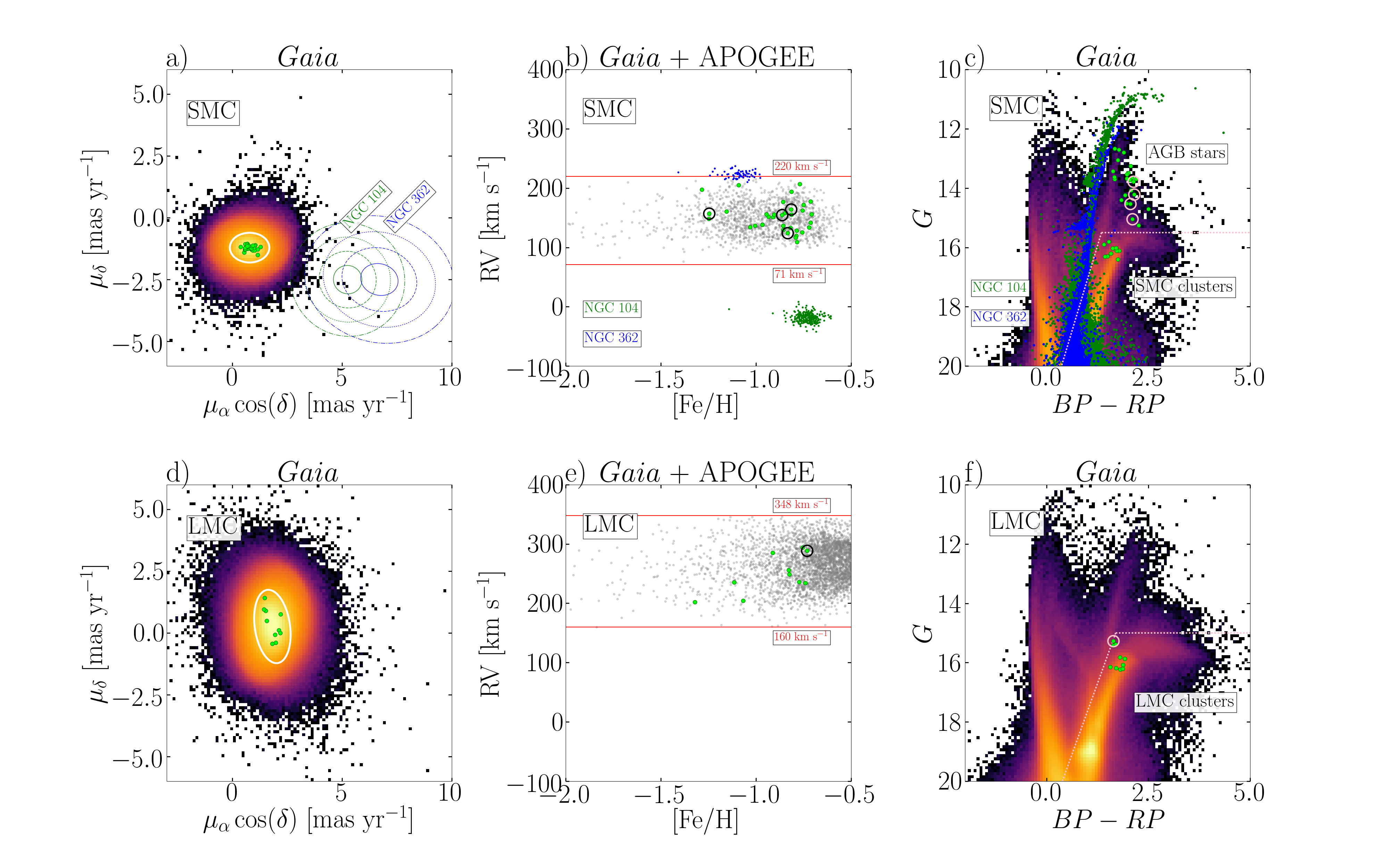}\\
		\caption{Panels (a) and (d): The proper motion plane (mas/yr) for the selected SMC, NGC 104, and NGC 362 (\textit{top}), and LMC (\textit{bottom}) candidates from \citet{Helmi2018}. The blue and green concentric ellipses show the 1$\sigma$, 2$\sigma$, 3$\sigma$ and 4$\sigma$ levels, while the while ellipses are the same defined in \citet{Nidever2020}. The lime dots in all the panels refer to the N-rich stars, with black open circles indicating the confirmed semi-regular (SR) variables \citep{Jayasinghe2020}. Panels (b) and (e): Radial velocity-metallicity plot for SMC/LMC candidate stars cross-identified in APOGEE-2S$+$ (gray dots). The red lines are the same as defined in \citet{Nidever2020}. Panels (c) and (f): The density distribution of stars in the \textit{Gaia} DR2 Color-Magnitude Diagram (CMD). The pink open circles indicate the location of the SR variables, while the pink dashed line roughly separates the region dominated by Magellanic star clusters and the possible location of evolved AGB stars.}
		\label{fig2}
	\end{center}
\end{figure*}

\begin{figure*}[t]
	\begin{center}
		\includegraphics[height = 13. cm]{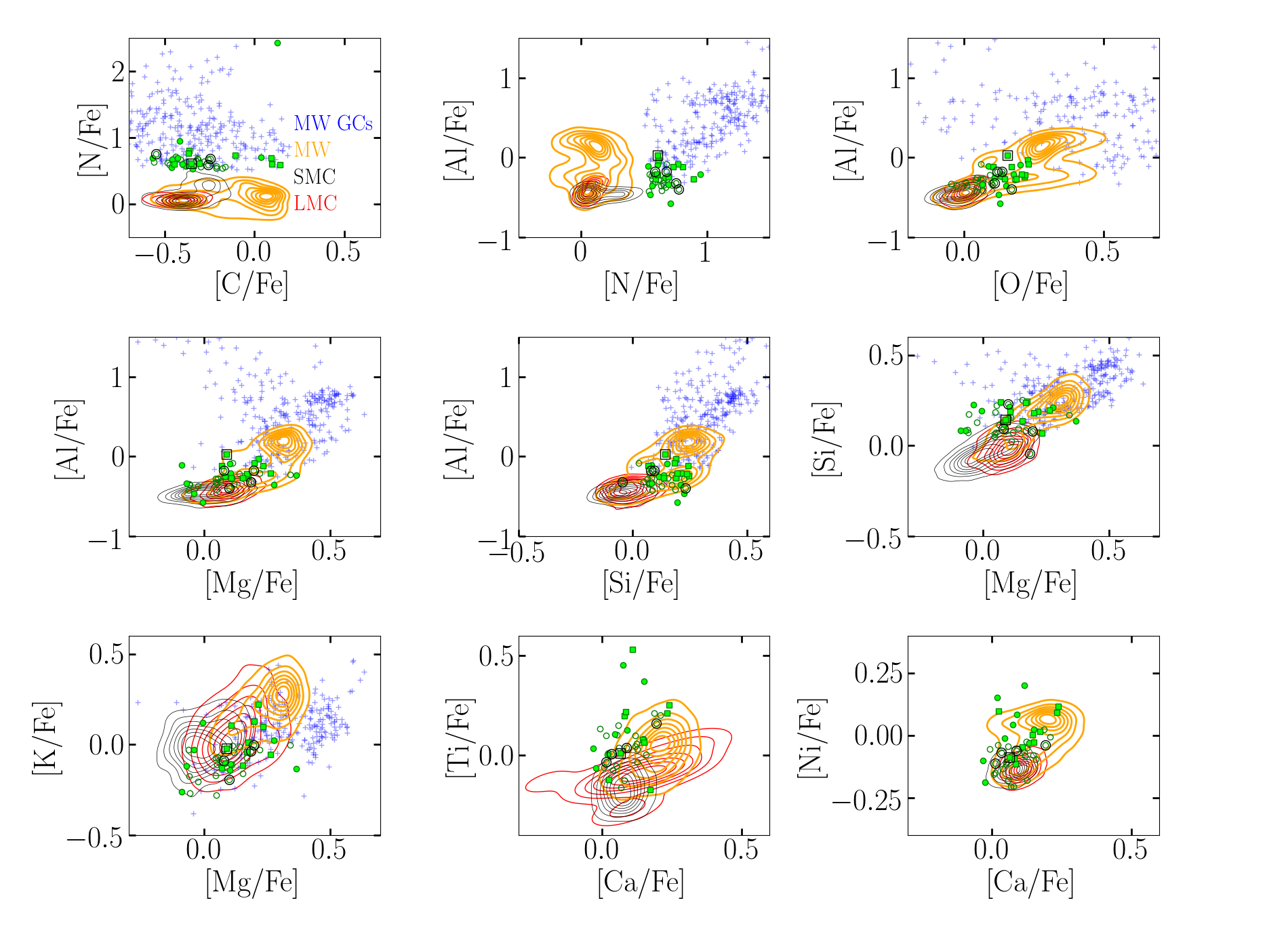}
		\caption{Distributions of various elemental-abundance ratios from the APOGEE-2S$+$ survey, represented by iso-abundance contours for the MW stars (orange), SMC (black), and  LMC (red) stars, and Galactic GCs (blue crosses) from \citet{Szabolcs2020}. Bright N-rich RGB stars in the SMC and LMC are marked as lime circles and squares, respectively, while the green unfilled circles indicate N-rich AGB stars in the SMC. The SR variables in the SMC and LMC are highlighted with black circles and squares, respectively. }
		\label{fig3}
	\end{center}
\end{figure*}
	
\begin{figure*}[t]
	\begin{center}
		\includegraphics[height = 17. cm]{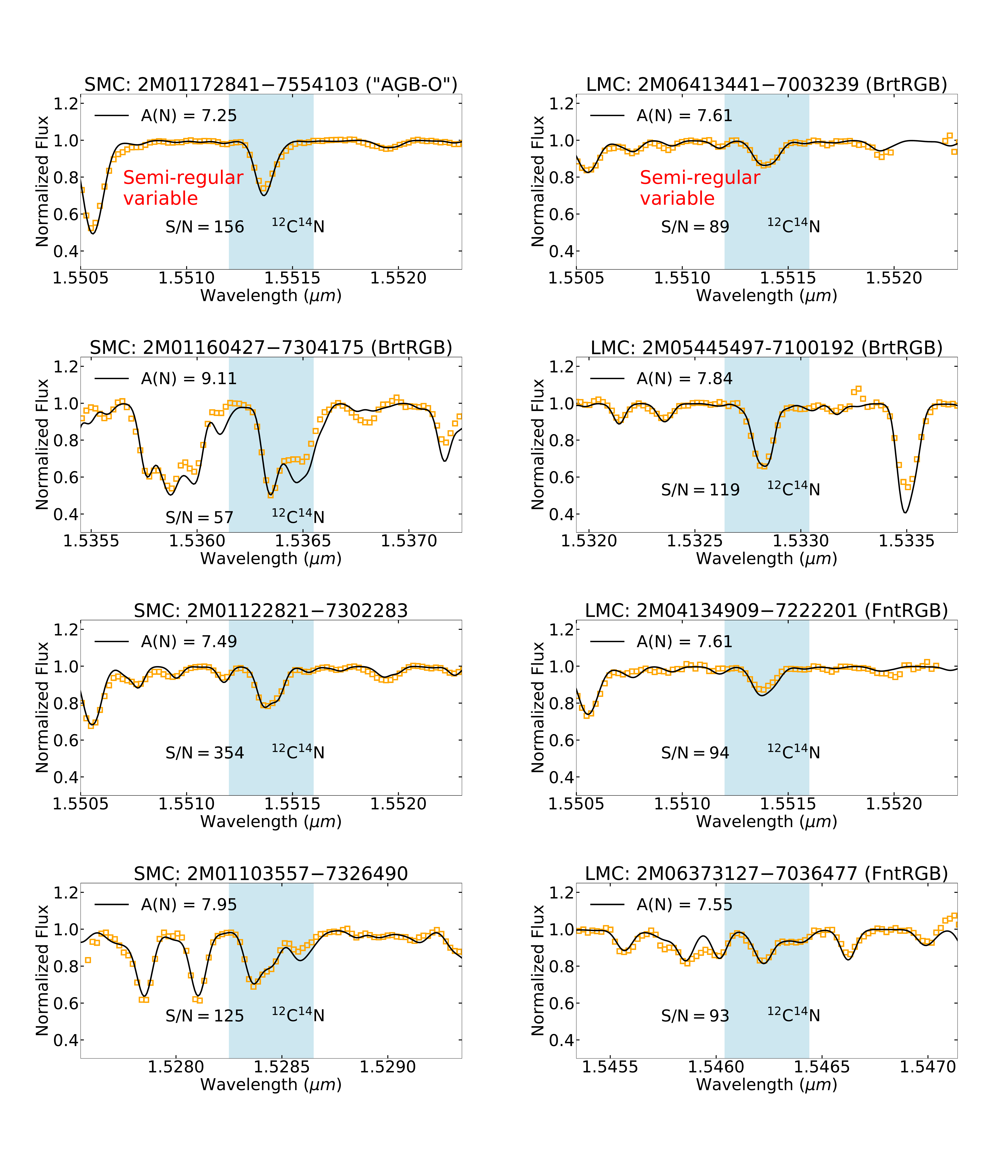}
		\caption{High-resolution near-IR \textit{H}-band spectrum of the newly identified N-rich stars in the Small (\textit{left}) and Large (\textit{right}) Magellanic Clouds, covering spectral regions around the $^{12}$C$^{14}$N band (orange squares). Superimposed is the best-fit of a MARCS/\texttt{BACCHUS} spectral synthesis (black line). The legends in each panel show the absolute abundance, $A$(N), and the signal-to-noise (S/N) in the region of the feature, respectively.}
		\label{fig4}
	\end{center}
\end{figure*}	
	
\section{DATA}
\label{section2}

In this work, we manually re-examined the high-resolution (\textit{R}$\sim$22,500) spectra in the \textit{H}-band ($\lambda\sim$1.5--1.7$\mu$m) from the APOGEE instrument \citep{Gunn2006, Wilson2019} that operates on the Ir\'en\'ee Du Pont 2.5m telescope \citep{Bowen1973} at Las Campanas Observatory (LCO, APOGEE-2S) as part of the incremental 16th data release of SDSS-IV \citep{Blanton2017, Ahumada2020}. These spectra include internal data through March 2020 from the APOGEE-2 survey \citep[][hereafter APOGEE-2S$+$]{Majewski2017} toward the Southern Hemisphere. Targeting strategies for APOGEE-2$+$, data reduction of the spectra, determination of radial velocities, and atmospheric parameters are fully described in \citet{Holtzman2015}, \citet{Nidever2015}, \citet{Ana2016}, and \citet{Zasowski2017}. All the spectra analyzed in this work are in the following ranges: \textit{(i)}  S/N $>$ 60 pixel$^{-1}$; \textit{(ii)} 3200 K $<$T$_{\rm eff}$ $<$ 5500 K; $-0.5 <$ $\log$ \textit{g}$< 5.5$; \textit{g}; (\textit{iii}) \texttt{ASPCAPFLAG} $== 0$.

We identified potential GC debris over $\sim$3,535 mildly metal-poor ($-2.0 \lsim$ [Fe/H] $\lsim-0.7$) stars toward the MCs in the [N/Fe]--[Fe/H], plane following the same methodology as described in \citet{Fernandez-Trincado2019}. This yielded the serendipitous discovery of 44 stars toward the MCs with stellar atmosphere strongly enriched in nitrogen, which are typically found among the so--called \textit{second-generation} of stars in Galactic GCs. The newly identified N-rich stars are shown as lime open symbols in Figure \ref{fig1}.

Although a handful of N-rich star candidates have been identified in the APOGEE survey toward the inner halo \citep{Martell2016} and Galactic Bulge \citep[][]{Schiavon2017}, based on the \texttt{ASPCAP}/APOGEE DR12 catalog, the \texttt{ASPCAP} pipeline \citep{Ana2016} contains some caveats that hamper a profound exploration and characterization of the mildly metal-poor N-rich population \citep[for a review, see][]{Fernandez-Trincado2019}. 

Here, we perform a spectral synthesis analysis, independently of the \texttt{ASPCAP} pipeline, to disentangle the underlying C, N, and O abundances from the $^{12}$C$^{16}$O, $^{12}$C$^{14}$N, and $^{16}$OH band strengths. To this purpose, we performed an LTE analysis with a MARCS grid of spherical models with the \texttt{BACCHUS} code \citep{Masseron2016}, adopting the same methodology as described in \citet{Fernandez-Trincado2016, Fernandez-Trincado2017, Fernandez-Trincado2019, Fernandez-Trincado2019a, Fernandez-Trincado2019b, Fernandez-Trincado2019c}. It is important to keep in mind that \texttt{ASPCAP} uses a global fit to the continuum in three detector chips independently, while we place the pseudo-continuum in a region around the lines of interest. We believe that our manual method is more reliable, since it avoids possible shifts in the continuum location due to imperfections in the spectral subtraction along the full spectral range. 

In order to provide a consistent chemical analysis, we re-determine the chemical abundances by means of a careful line selection, and measure abundances based on a line-by-line basis with the \texttt{BACCHUS} code, and by adopting the line selection for the various elements as in \citet{Fernandez-Trincado2019}. Finally, we re-derived chemical abundances adopting as input the uncalibrated effective temperatures (T$^{\rm unc}_{\rm eff}$), surface gravities $\log{}$ \textit{g}$^{\rm unc}$ and the overall metallicity ([M/Fe]) from the \texttt{ASPCAP} pipeline. We do not calculate chemical abundances based on the photometric atmospheric parameters, as they become error dominated for stars toward the MCs, due the difficulty of calculating accurate reddenings in these regions \citep[see, e.g.,][]{Nidever2020}, making them unsuitable to estimate precise chemical abundances for stars in the inner MCs. 

\section{Results and Analysis}
\label{section3}

We find that the newly identified N-rich stars span a wide range of metallicities ($-1.4 <$[Fe/H]$<-0.7$), peaking at [Fe/H]$\sim-0.89$, and that they exhibit nitrogen abundances well above typical Galactic levels over a range of metallicities, which is $\gtrsim$3--4$\sigma$ above the typical MW [N/Fe]. It seems likely that whatever nucleosynthesis process is responsible for these nitrogen over-abundances in the field of the MCs is similar to that which caused the unusual stellar populations in Galactic GCs at similar metallicity.

Figures \ref{fig1}a--d reveal the existence of a large number of N-rich stars toward the MCs. There are 34 out of 44 N-rich stars located at the center of the SMC, whereas ten stars in our sample reside on the periphery of the LMC.

Figures \ref{fig2}a;d show the proper motions of \textit{Gaia} DR2, confirming that the N-rich stars are permanent residents of the MCs. In the case of N-rich stars in the SMC, it is clearly visible that its proper motion distribution deviates by more than 4$\sigma$ from the nominal proper motions of the NGC 104 and NGC 362 GCs. Figures \ref{fig2}b;e also reveal that the radial-velocity distributions differ from those of the field GCs, while the Color-Magnitude Diagram (CMD) using \textit{Gaia} bands (see Figures \ref{fig2}c;f) also rules out the possibility these N-rich stars are stellar debris of these two Galactic GCs. This is also supported by inspection of [Fe/H]; the N-rich stars exhibit a larger metallicity scatter, on average being more metal rich than NGC 362 and more metal poor than NGC 104. Based on these properties, we conclude that these are \textit{bona fide} N-rich stars in the MCs, which are chemically identical to those identified toward the bulge and halo of the MW \citep[see, e.g.,][]{Fernandez-Trincado2017, Fernandez-Trincado2019, Fernandez-Trincado2019b, Fernandez-Trincado2019c}. 

Figure \ref{fig3} presents some of the light-, odd-Z, and $\alpha$-element patterns, and demonstrates a very distinct separation in nitrogen, with a star-to-star scatter $\Delta_{\rm [N/Fe]} \gtrsim +0.28$ dex and $\Delta_{\rm [C/Fe]} \gtrsim +0.25$ dex, which is moderately anti-correlated with [C/Fe], and runs roughly between \{[C/Fe], [N/Fe]\}$=$\{$-0.6$, $+0.53$\} and \{$+0.15$, $+1.0$\}. We have found mean values for [C/Fe], [N/Fe], [O/Fe], [Mg/Fe], [Al/Fe], [Si/Fe], [K/Fe], [Ce/Fe], and [Nd/Fe] that are compatible with Galactic GC stars \citep{Szabolcs2020} at similar metallicity. There is also a star with [N/Fe]$\gtrsim+2.42$ that exceeds the extreme abundance patterns seen in Galactic GCs, highlighting the uniqueness of these stars. Overall, we can see clearly from Figure \ref{fig3} that our sample of stars in the MCs behave in a similar way as MW GC stars, supporting the idea that most of the newly identified stars could be related to MC GCs.   

The newly identified N-rich population is separated relatively cleanly from MW stars and MC stars in the [C/Fe] vs. [N/Fe] and [N/Fe] vs. [Al/Fe] planes. In general, these N-rich stars exhibit slightly higher abundance ratios in Al, Si, Ti, and Ni compared to the SMC and LMC populations, but they exhibit lower abundance ratios in O, Mg, Al, Si, K, and C compared to MW field stars, with the $\alpha-$elements (O, Mg, Si, and Ca) at comparable levels ($\sim +0.1$) to low-$\alpha$ halo MW field stars \citep[e.g.,][]{Hayes2018}. 

The star-to-star scatter is between 0.1 -- 0.3 dex for the different chemical species, being slightly lower for the $\alpha-$elements. Therefore, the star-to-star scatter in iron and other chemical species could be attributed to different progenitors, which could explain the observed chemical anomalies toward the MCs, in a similar manner as observed in and around the MW halo metal-poor stars today. The mildly metal-poor N-rich stars toward the MCs may have formed following minor merger events in the early history of the MCs. However, there are other observational features in our sample that allow us to invoke other possible scenarios to explain the observed abundance patterns toward the MCs.

Our sample includes five SR variable stars reported in the ASAS-SN Catalog of Variable Stars \citep{Jayasinghe2020}. Four of them are located toward the SMC, with one been selected as a possible O-rich AGB star based on its position in the (J-Ks, H) diagram \citep[see][]{Nidever2020}. We also identified a SR variable in our sample toward the LMC, selected as a bright RGB in \citet{Nidever2020}. These SR stars have periods between $\sim$85 and 757 days and variability amplitudes between 0.17 and 0.51 mag in the V-band. In conclusion, we find evidence that the SR stars are neither carbon rich nor oxygen rich, but exhibit lower carbon and oxygen abundance ratios, [C/Fe]$<+0.15$ and [O/Fe]$\lesssim +0.23$. It is worthing to mention that no bias or uncertainties are introduced in our spectroscopic analysis, as is the case for, e.g., shorter-period Cepheids or RR Lyrae stars \citep{Pancino2015}.

Thus, the observed nitrogen over-abundances and the modest enhancements of the \textit{s}-process elements, coupled with the apparent variability of these SR stars, suggest that some of the evolved objects could be likely intermediate-mass ($\sim$3--5M$_{\odot}$) AGB stars (one of the likely agent that self-enriches the GCs) that have undergone hot bottom burning and are becoming N-rich, according to chemical evolution models \citep{Karakas2018}, but without production of significant amounts of aluminum, as envisioned by \citet{Ventura2016}. These SR N-rich stars have remarkably stronger $^{12}$C$^{14}$N lines (see Figure \ref{fig4}) compared to other stars with similar relevant parameters; it can be asserted that these have much higher nitrogen abundance. The presence of such young, mildly metal-poor stellar populations in the MCs has important implications. Thus, the interpretation of our results depends crucially on establishing the evolutionary stage of the stars under analysis. It is also important to note that there are no known MC GCs within an angular separation of approximately one arcmin of these stars.

In this context, one can immediately notice that two sub-samples occupy different loci in the CMD displayed in Figure \ref{fig2}c;f. N-rich stars with \textit{G} $\lesssim 15.0$ (LMC) and \textit{G} $\lesssim 15.5$ (SMC) occupy the same locus as MC AGB stars (referred to as "N-rich AGB" stars henceforth), while the N-rich stars with fainter \textit{G} magnitudes roughly occupy the same locus as the bright RGB stars in the MC stellar clusters. These stars are tagged as genuine \textit{migrants} from MC GCs (hereafter N-rich BrRGB stars), and are among the oldest objects in the MCs. The existence of N-rich AGB stars in our sample can be also further assessed by the possible presence of circumstellar dust \citep{Habing1996}, as the N-rich AGB stars occupy a locus towards colors that are redder than those N-rich BrRGB stars in the CMD diagram. In particular, 55\% of the N-rich stars in our sample inhabit the AGB part of the diagram, which provides further evidence for an important contribution of AGB stars to our N-rich sample. 

Although these stars have elemental abundances consistent with each other, we find that in the N-rich AGB stars the oxygen abundance ratios -- generally lower than the [O/Fe] of N-rich BrRGB stars--this lends further support to the notion that the two populations do not share the same origin.

It is also interesting to note that all the N-rich AGB stars in our sample were identified toward the SMC system, while the N-rich BrRGB stars are present in both, and likely could be part of the oldest stars in the MCs. We also conclude that there is significant evidence for a large contribution of possible AGB stars to our sample toward the MCs, suggesting that the detection of N-rich stars toward the MW has been biased towards RGB stars \citep[e.g.,][]{Schiavon2017}, which should result in a substantial difference in AGB contribution to the N-rich sample and the rest of the field, as already noted in \citet{Fernandez-Trincado2019}. 

Our finding can be understood in terms of different scenarios. Here, we conjecture that there may be at least two possible channels for the production of N-rich stars in the MCs: (\textit{i}) the N-rich BrRGB stars could be former members of a population of GCs that was previously dissolved and/or evaporated in the LMC/SMC, and were later incorporated into the field of the MCs themselves--'smoking gun' evidence that they have been accreted along with their now-disrupted host GCs. Such a scenario could potentially explain the predominance of N-rich BrRGB stars that are currently not gravitationaly bound to any MC clusters. The chemical patterns of these stars are identical or comparable to those seen in old MW GC stars \citep[e.g.,][]{Szabolcs2020}, and possibly associated with MC GCs, with ages between $\sim$2 and 10 Gyr \citep{Hollyhead2018, Lagioia2019, Milone2020}. In support of this scenario, one would expect to find N-rich BrRGB stars in the same environments as GCs today, as at least some of them would have been formed in the same molecular clouds as the GCs themselves. Thus, this observational finding would suggest that either some old GCs in the MCs have possibly experienced significant stellar mass-loss \citep[e.g.,][]{Mackey2007, Dalessandro2016}. This would suggest a common, single pathway for the formation and evolution of old ($\gtrsim 2$ Gyr) GCs within the Local Group \citep{Martocchia2017}; (\textit{ii}) On the other hand, the discovery of a significant population of N-rich stars in the AGB evolutionary stage (and possibly of intermediate-mass), further supports the idea that AGB stars are possibly one of the key players in the pollution of the intracluster medium \citep[e.g.,][]{Ventura2016} proposed to explain the formation of MPs in GCs. 

The presence of such a significant young and moderately metal-poor stellar populations in the MCs would have interesting consequences for the understanding of the formation and evolution of GC systems in the Local Universe, i.e.,  the presence of star-to-star abundance spreads in this possible "young" N-rich AGB population appears to be at odds with the apparent near-exclusivity of this population within old GCs. 

\section{Concluding remarks}
\label{section4} 
 
We report the serendipitous discovery of a large population of mildly metal-poor N-rich stars toward the LMC/SMC. Our sample is composed mainly of stars in the bright RGB (45\%) and AGB (55\%) evolutionary stage. This sample adds to the literature nitrogen measurements for several new stars; to our knowledge, none of the large spectroscopic surveys of the LMC/SMC have so far included measurements of significant nitrogen abundances. 

The discovery of two sub-populations of N-rich stars suggests that the occurrence of chemical anomalies (also crucial in the chemical makeup of MPs in all GCs) might not be exotic events from the past, but can also form at lower redshift, as also envisaged by \citet{Bekki2019}, and not limited to stars with masses less than $\sim$1.6 M$_{\odot}$ \citep[see, e.g.,][]{Bastian2018}.

Our findings motivate the future search for N-rich stars in Local Group dwarf galaxies that could have once hosted GCs, now in the form of disrupted remnants. As light-element variations have, however, not been found amongst the dwarf's field stars \citep{Geisler2007, Villanova2017}.

\acknowledgments

We thank the referee for insightful comments that improved the paper. J.G.F-T. is supported by FONDECYT No. 3180210. T.C.B. and V.M.P. acknowledge partial support for this work from grant PHY 14-30152; Physics Frontier Center / JINA Center for the Evolution of the Elements (JINA-CEE), awarded by the US National Science Foundation. S.V. gratefully acknowledges the support provided by Fondecyt regular n. 1170518. AR-L acknowledges financial support provided in Chile by Comisi\'on Nacional de Investigaci\'on Cient\'ifica y Tecnol\'ogica (CONICYT) through the FONDECYT project 1170476 and by the QUIMAL project 130001. B.B. acknowledge partial financial support from the brazilian agencies CAPES - Financial code 001, CNPq and FAPESP. L.C. thanks Gloria Delgado-Inglada for useful discussions. We also acknowledge the support of Universidad de Concepci\'on for providing HPC resources on the Supercomputer TITAN.  

This work has made use of data from the European Space Agency (ESA) mission Gaia (\url{http://www.cosmos.esa.int/gaia}), processed by the Gaia Data Processing and Analysis Consortium (DPAC, \url{http://www.cosmos.esa.int/web/gaia/dpac/consortium}). Funding for the DPAC has been provided by national institutions, in particular the institutions participating in the Gaia Multilateral Agreement.

Funding for the Sloan Digital Sky Survey IV has been provided by the Alfred P. Sloan Foundation, the U.S. Department of Energy Office of Science, and the Participating Institutions. SDSS- IV acknowledges support and resources from the Center for High-Performance Computing at the University of Utah. The SDSS web site is www.sdss.org.

SDSS-IV is managed by the Astrophysical Research Consortium for the Participating Institutions of the SDSS Collaboration including the Brazilian Participation Group, the Carnegie Institution for Science, Carnegie Mellon University, the Chilean Participation Group, the French Participation Group, Harvard-Smithsonian Center for Astrophysics, Instituto de Astrof\`{i}sica de Canarias, The Johns Hopkins University, Kavli Institute for the Physics and Mathematics of the Universe (IPMU) / University of Tokyo, Lawrence Berkeley National Laboratory, Leibniz Institut f\"{u}r Astrophysik Potsdam (AIP), Max-Planck-Institut f\"{u}r Astronomie (MPIA Heidelberg), Max-Planck-Institut f\"{u}r Astrophysik (MPA Garching), Max-Planck-Institut f\"{u}r Extraterrestrische Physik (MPE), National Astronomical Observatory of China, New Mexico State University, New York University, the University of Notre Dame, Observat\'{o}rio Nacional / MCTI, The Ohio State University, Pennsylvania State University, Shanghai Astronomical Observatory, United Kingdom Participation Group, Universidad Nacional Aut\'{o}noma de M\'{e}xico, University of Arizona, University of Colorado Boulder, University of Oxford, University of Portsmouth, University of Utah, University of Virginia, University of Washington, University of Wisconsin, Vanderbilt University, and Yale University.

\floattable
\begin{deluxetable*}{ccccccccccccccccccccccccc}
	\center
	\setlength{\tabcolsep}{0.7mm}  
	\tabletypesize{\tiny}
	\tablecolumns{25}
	\rotate
	\tablecaption{Basic parameters and Abundances of the MC N-rich stars.}
	\tablehead{
		\colhead{APOGEE$-$Id}		    &  
		\colhead{Target Class}		    &        
		\colhead{\textit{G}-band}		    &        
		\colhead{$\#$Visits}		    & 
		\colhead{S/N}		    &  
		\colhead{RV}		    &    
		\colhead{$\sigma$RV}		    &        
		\colhead{T$_{\rm eff}$}		   & 
		\colhead{$\log$ \textit{g}}		     & 
		\colhead{ $\xi_{t}$}		    & 
		\colhead{[M/H]}		          & 
		\colhead{[C/Fe]}		    &        
		\colhead{[N/Fe]}		    & 
		\colhead{[O/Fe]}		    & 
		\colhead{[Mg/Fe]}		    &   
		\colhead{[Al/Fe]}		    &  
		\colhead{[Si/Fe]}		    &    
		\colhead{[K/Fe]}		    &        
		\colhead{[Ca/Fe]}		    & 
		\colhead{[Ti/Fe]}		    & 
		\colhead{[Fe/H]}		    & 
		\colhead{[Ni/Fe]}		    & 		
		\colhead{[Ce/Fe]}		    &        
		\colhead{[Nd/Fe]}		    & 
		\colhead{[Yb/Fe]}		\\
		\colhead{}		    &  
		\colhead{}		    &      
		\colhead{mag}		    &          
		\colhead{}		    & 
		\colhead{pixel$^{-1}$}		    &  
		\colhead{km s$^{-1}$}		    &    
		\colhead{km s$^{-1}$}		    &        
		\colhead{K}		   & 
		\colhead{dex}		     & 
		\colhead{km s$^{-1}$}		    & 
		\colhead{}		          & 
		\colhead{}		    &        
		\colhead{}		    &        
		\colhead{}		    &        
		\colhead{}		    &        
		\colhead{}		    &        
		\colhead{}		    &        
		\colhead{}		    &        
		\colhead{}		    &        
		\colhead{}		    &        
		\colhead{}		    &        
		\colhead{}		    &        
		\colhead{}		    &        
		\colhead{}		    &        
		\colhead{}		    
	}
	\startdata
2M05445497$-$7100192  & LMC/BrtRGB & 16.08 & 9  &  132  & 234.4 & 0.4 & 3786 &    0.28 & 2.81 & $-$0.74 & $-$0.36 & $+$0.80 & $+$0.17 & $+$0.23 & $-$0.11 & $+$0.06 & $+$0.09 & $+$0.14 & $+$0.07 & $-$0.89 & $+$0.01 & $+$0.31 & $+$0.55 & $+$0.72 \\
2M05222260$-$7239312  & LMC/BrtRGB & 15.87 & 9  &  127  & 236.2 & 0.3 & 3833 &    0.44 & 2.88 & $-$0.78 & $-$0.30 & $+$0.61 & $+$0.22 & $+$0.21 & $-$0.02 & $+$0.18 & $+$0.22 & $+$0.23 & $+$0.21 & $-$0.99 & $+$0.09 & $+$0.33 & $+$0.56 & $+$0.69 \\
2M05113944$-$6619050  & LMC/BrtRGB & 15.83 & 12 &  138  & 285.0 & 2.5 & 3832 &    0.41 & 2.50 & $-$0.93 & $+$0.09 & $+$0.69 & $+$0.21 & $+$0.17 & $-$0.24 & $+$0.24 & $+$0.01 & $+$0.07 & $+$0.19 & $-$0.94 & $-$0.09 & $+$0.19 & $+$0.27 &    ...  \\
2M06373127$-$7036477  & LMC/FntRGB & 16.18 & 9  &   74  & 255.9 & 0.3 & 3952 &    0.65 & 2.82 & $-$0.84 & $+$0.14 & $+$0.59 & $+$0.18 & $+$0.10 & $-$0.27 & $+$0.19 & $+$0.10 & $+$0.02 & $-$0.12 & $-$0.89 & $+$0.10 & $+$0.18 & $+$0.33 & $+$0.57 \\
2M06413441$-$7003239  & LMC/BrtRGB & 15.27 & 11 &  129  & 288.9 & 0.3 & 4047 &    0.88 & 2.53 & $-$0.74 & $-$0.35 & $+$0.60 & $+$0.15 & $+$0.08 & $+$0.02 & $+$0.14 & $-$0.02 & $+$0.06 & $+$0.01 & $-$0.77 & $+$0.08 & $+$0.35 & $+$0.33 & $+$0.64 \\
2M04412976$-$7439249  & LMC/FntRGB & 16.23 & 14 &  107  & 248.6 & 5.5 & 3840 &    0.55 & 2.57 & $-$0.83 & $+$0.09 & $+$0.60 & $+$0.22 & $+$0.16 & $-$0.21 & $+$0.23 & $-$0.03 & $+$0.17 & $-$0.17 & $-$0.82 & $+$0.02 & $+$0.15 & $+$0.31 & $+$0.58 \\
2M04420596$-$7600190  & LMC/BrtRGB & 15.43 & 14 &  167  & 202.1 & 0.2 & 4081 &    0.86 & 2.42 & $-$1.33 & $-$0.44 & $+$0.59 & $+$0.20 & $+$0.26 & $-$0.21 & $+$0.19 & $-$0.05 & $+$0.10 & $+$0.53 & $-$1.29 & $-$0.05 & $+$0.17 &    ...  &    ...  \\
2M04130396$-$7242541  & LMC/BrtRGB & 15.38 & 11 &  158  & 204.2 & 0.2 & 4026 &    0.31 & 2.97 & $-$1.06 & $-$1.05 & $+$0.89 & $+$0.14 & $+$0.15 & $-$0.27 & $+$0.13 & $-$0.11 & $+$0.15 & $+$0.06 & $-$1.05 & $-$0.03 & $+$0.35 & $+$0.83 & $+$0.98 \\
2M04134909$-$7222201  & LMC/FntRGB & 16.21 & 11 &  106  & 235.6 & 0.6 & 3901 &    0.65 & 2.96 & $-$1.13 & $-$0.36 & $+$0.76 & $+$0.23 & $+$0.19 & $-$0.08 & $+$0.18 & $+$0.12 & $+$0.23 & $+$0.25 & $-$1.22 & $+$0.12 & $+$0.19 & $+$0.42 &    ...  \\
2M06075280$-$6417424  & LMC/...    & 16.15 & 9  &   81  & 293.9 & 7.8 & 4069 &    0.96 & 1.86 & $-$0.76 & $-$0.10 & $+$0.73 & $+$0.10 & $+$0.06 & $-$0.11 & $+$0.24 & $-$0.12 & $+$0.08 & $+$0.21 & $-$0.74 & $-$0.11 & $+$0.40 & $+$0.32 & $+$1.05 \\
\hline
2M00503421$-$7303142  & SMC/AGB-O  & 15.26 & 4  &  145  & 133.1 & 0.6 & 3652 &    0.03 & 2.41 & $-$0.83 & $-$0.21 & $+$0.53 & $+$0.04 & $-$0.06 & $-$0.42 & $+$0.08 & $-$0.03 & $+$0.12 & $-$0.03 & $-$0.72 & $-$0.18 & $+$0.23 &    ...  &    ...  \\
2M00531362$-$7251451  & SMC/AGB-O  & 13.69 & 4  &  269  & 155.9 & 0.8 & 3759 & $-$0.02 & 2.91 & $-$0.67 & $-$0.25 & $+$0.67 & $+$0.06 & $+$0.11 & $-$0.07 & $+$0.02 & $-$0.00 & $+$0.14 & $+$0.08 & $-$0.74 & $-$0.08 & $+$0.36 & $+$0.76 & $+$0.65 \\
2M01015101$-$7133002  & SMC/BrtRGB & 16.11 & 15 &  107  & 134.0 & 0.8 & 3893 &    0.83 & 2.81 & $-$1.05 & $-$0.38 & $+$0.56 & $+$0.15 & $+$0.27 & $-$0.35 & $+$0.20 & $+$0.02 & $+$0.07 & $+$0.45 & $-$1.10 & $+$0.04 & $+$0.39 & $+$0.86 & $+$0.65 \\
2M01124858$-$7251211  & SMC/FntRGB & 16.11 & 15 &  121  & 162.6 & 0.3 & 4017 &    1.01 & 2.99 & $-$0.75 & $-$0.45 & $+$0.66 & $+$0.11 & $+$0.08 & $-$0.25 & $+$0.11 & $-$0.10 & $+$0.08 & $+$0.12 & $-$0.79 & $+$0.08 & $+$0.37 & $+$0.66 & $+$0.97 \\
2M01131142$-$7338425  & SMC/FntRGB & 16.02 & 12 &  107  & 109.0 & 0.5 & 4082 &    1.33 & 2.34 & $-$0.79 & $-$0.56 & $+$0.69 & $+$0.06 & $-$0.07 & $-$0.33 & $+$0.08 & $-$0.11 & $-$0.03 & $+$0.03 & $-$0.75 & $-$0.10 & $+$0.37 & $+$0.65 &    ...  \\
2M01160427$-$7304175  & SMC/BrtRGB & 15.81 & 15 &  129  & 148.3 & 0.3 & 3863 &    0.21 & 2.07 & $-$1.24 & $+$0.12 & $+$2.42 & $-$0.01 &    ...  &    ...  & $-$0.02 &    ...  &    ...  & $-$0.39 & $-$1.27 & ...     & $+$0.92 & $+$1.01 &    ...  \\
2M01050424$-$7545352  & SMC/FntRGB & 16.30 & 18 &  104  & 197.5 & 3.5 & 4201 &    1.12 & 2.19 & $-$1.29 & $-$0.41 & $+$0.95 & $+$0.17 & $+$0.10 & $-$0.21 & $+$0.21 & $-$0.10 & $+$0.11 &    ...  & $-$1.30 & $+$0.20 &    ...  &    ...  &    ...  \\
2M01081022$-$7448592  & SMC/FntRGB & 16.41 & 18 &  100  & 205.2 & 0.6 & 3920 &    0.61 & 2.99 & $-$1.10 & $-$0.42 & $+$0.69 & $+$0.13 & $+$0.36 & $-$0.23 & $+$0.13 & $-$0.13 & $+$0.15 & $+$0.37 & $-$1.17 & $+$0.03 & $+$0.22 & $+$1.00 & $+$1.04 \\
2M01125370$-$7455328  & SMC/FntRGB & 16.31 & 17 &   69  & 150.9 & 0.5 & 4359 &    1.55 & 1.59 & $-$0.89 & $-$0.40 & $+$0.60 & $+$0.20 & $+$0.10 & $-$0.09 & $+$0.15 & $-$0.14 & $+$0.04 & $+$0.10 & $-$0.83 & $-$0.01 &    ...  &    ...  & $+$0.88 \\
2M01172841$-$7554103  & SMC/AGB-O  & 15.05 & 18 &  271  & 156.9 & 2.9 & 3663 & $-$0.13 & 2.12 & $-$1.27 & $-$0.76 & $+$0.77 & $+$0.17 & $+$0.09 & $-$0.39 & $+$0.23 & $-$0.19 & $+$0.01 & $-$0.03 & $-$1.13 & $-$0.11 & $+$0.09 & $+$0.12 &    ...  \\
2M01173245$-$7509207  & SMC/AGB-O  & 14.52 & 17 &  323  & 151.9 & 1.4 & 3741 &    0.16 & 2.89 & $-$0.96 & $-$0.48 & $+$0.62 & $+$0.18 & $+$0.05 & $-$0.27 & $+$0.22 & $-$0.06 & $+$0.06 & $-$0.16 & $-$0.85 & $-$0.20 & $+$0.16 & $+$0.32 &    ...  \\
2M01354128$-$7131328  & SMC/FntRGB & 16.20 & 15 &  105  & 136.6 & 0.4 & 4161 &    1.41 & 2.09 & $-$1.02 & $-$0.46 & $+$0.54 & $+$0.12 & $-$0.04 & $-$0.46 & $+$0.22 & $-$0.03 & $-$0.02 & $-$0.06 & $-$0.93 & $-$0.19 & $+$0.29 &    ...  & $+$0.91 \\
2M01384622$-$7127456  & SMC/BrtRGB & 16.02 & 15 &  131  & 160.7 & 0.5 & 3894 & $-$0.06 & 2.76 & $-$1.17 & $+$0.03 & $+$0.70 & $+$0.12 & $-$0.00 & $-$0.57 & $+$0.19 & $+$0.12 & $+$0.13 &    ...  & $-$1.23 & $-$0.03 & $+$0.39 &    ...  &    ...  \\
2M01440256$-$7136229  & SMC/FntRGB & 15.89 & 15 &  108  & 155.6 & 0.3 & 4244 &    1.61 & 1.66 & $-$0.82 & $-$0.34 & $+$0.53 & $+$0.08 & $-$0.08 & $-$0.10 & $+$0.08 & $-$0.25 & $+$0.01 & $+$0.01 & $-$0.73 & $+$0.15 &    ...  & $+$0.34 &    ...  \\
2M00381852$-$7306415  & SMC/...    & 14.70 & 1  &  104  & 156.0 & ... & 3648 & $-$0.06 & 2.99 & $-$0.93 & $-$1.19 & $+$1.05 & $+$0.20 & $+$0.15 &    ...  & $+$0.25 & $-$0.15 & $+$0.18 & $+$0.21 & $-$0.89 & $+$0.03 & $+$0.24 & $+$0.29 &    ...  \\
2M01053079$-$7139594  & SMC/...    & 13.65 & 1  &   94  & 136.6 & ... & 3645 & $-$0.03 & 2.75 & $-$0.85 & $-$0.37 & $+$0.63 & $+$0.07 & $+$0.08 &    ...  & $+$0.08 & $-$0.12 & $+$0.05 & $-$0.05 & $-$0.77 & $-$0.11 & $+$0.18 & $+$0.31 &    ...  \\
2M01021532$-$7259088  & SMC/...    & 14.21 & 1  &  113  & 164.1 & ... & 3675 & $-$0.02 & 2.69 & $-$0.80 & $-$0.24 & $+$0.67 & $+$0.13 & $+$0.19 & $-$0.17 & $+$0.07 & $-$0.00 & $+$0.19 & $+$0.16 & $-$0.92 & $-$0.04 & $+$0.29 & $+$0.37 &    ...  \\
2M00573209$-$7203203  & SMC/...    & 13.49 & 1  &  148  & 142.1 & ... & 3764 & $-$0.02 & 2.89 & $-$0.69 & $-$0.21 & $+$0.57 & $+$0.09 & $+$0.14 & $-$0.21 & $+$0.09 & $-$0.01 & $+$0.12 & $+$0.03 & $-$0.79 & $-$0.05 & $+$0.31 & $+$0.45 & $+$0.58 \\
2M00505642$-$7317378  & SMC/...    & 14.07 & 1  &   93  & 206.9 & ... & 3743 &    0.02 & 2.91 & $-$0.75 & $-$0.32 & $+$0.65 & $+$0.09 & $+$0.04 & $-$0.32 & $+$0.04 & $-$0.17 & $+$0.15 & $+$0.01 & $-$0.76 & $-$0.09 & $+$0.24 & $+$0.65 &    ...  \\
2M01015801$-$7241586  & SMC/...    & 13.73 & 1  &  130  & 177.7 & ... & 3755 &    0.05 & 2.97 & $-$0.70 & $-$0.16 & $+$0.56 & $+$0.13 & $+$0.34 & $-$0.22 & $+$0.17 & $-$0.00 & $+$0.20 & $+$0.20 & $-$0.87 & $-$0.01 & $+$0.24 &    ...  & $+$0.72 \\
2M00361550$-$7339480  & SMC/...    & 13.67 & 1  &  137  & 133.4 & ... & 3788 &    0.08 & 2.78 & $-$0.69 & $-$0.31 & $+$0.66 & $+$0.04 & $+$0.07 & $-$0.25 & $+$0.12 & $-$0.02 & $+$0.11 & $+$0.04 & $-$0.74 & $-$0.07 & $+$0.36 & $+$0.79 & $+$0.61 \\
2M01103557$-$7326490  & SMC/...    & 14.53 & 2  &  150  & 154.7 & 0.0 & 3726 &    0.11 & 2.99 & $-$0.88 & $-$0.54 & $+$0.75 & $+$0.11 & $+$0.18 & $-$0.31 & $-$0.04 & $-$0.03 & $+$0.03 & $+$0.00 & $-$0.83 & $-$0.07 & $+$0.17 & $+$0.23 &    ...  \\
2M01062078$-$7203288  & SMC/...    & 13.78 & 1  &  131  & 124.3 & ... & 3692 &    0.12 & 2.82 & $-$0.84 & $-$0.25 & $+$0.58 & $+$0.12 & $+$0.07 & $-$0.17 & $+$0.09 & $-$0.08 & $+$0.08 & $+$0.03 & $-$0.82 & $-$0.07 & $+$0.27 & $+$0.67 &    ...  \\
2M01020419$-$7208498  & SMC/...    & 12.79 & 1  &  195  & 119.4 & ... & 3866 &    0.13 & 2.76 & $-$0.77 & $-$0.31 & $+$0.68 & $+$0.06 & $+$0.16 & $-$0.27 & $+$0.08 & $-$0.07 & $+$0.05 & $+$0.10 & $-$0.86 & $-$0.11 & $+$0.36 &    ...  & $+$0.65 \\
2M00453676$-$7308450  & SMC/...    & 13.26 & 1  &  153  & 128.4 & ... & 3952 &    0.18 & 2.82 & $-$0.74 & $-$0.34 & $+$0.57 & $-$0.04 & $+$0.05 & $-$0.32 & $+$0.07 & $-$0.11 & $+$0.01 & $-$0.05 & $-$0.71 & $-$0.16 & $+$0.34 & $+$0.62 &    ...  \\
2M01011946$-$7206485  & SMC/...    & 12.70 & 1  &  193  & 155.3 & ... & 3980 &    0.21 & 2.89 & $-$0.88 & $-$0.47 & $+$0.58 & $+$0.00 & $+$0.06 & $-$0.31 & $+$0.11 & $-$0.10 & $+$0.03 & $-$0.07 & $-$0.80 & $-$0.13 & $+$0.28 & $+$0.66 & $+$0.88 \\
2M01122821$-$7302283  & SMC/...    & 14.24 & 2  &  151  & 194.2 & 0.1 & 3816 &    0.22 & 2.88 & $-$0.81 & $-$0.17 & $+$0.54 & $+$0.13 & $+$0.22 & $-$0.21 & $+$0.07 & $-$0.02 & $+$0.11 & $+$0.09 & $-$0.91 & $-$0.08 & $+$0.30 & $+$0.50 & $+$0.75 \\
2M01115889$-$7318168  & SMC/...    & 14.39 & 2  &  137  & 171.7 & 0.3 & 3855 &    0.28 & 2.94 & $-$0.74 & $-$0.32 & $+$0.65 & $+$0.10 & $+$0.11 & $-$0.34 & $+$0.07 & $-$0.04 & $+$0.12 & $+$0.05 & $-$0.76 & $-$0.02 &    ...  & $+$0.44 & $+$0.53 \\
2M00511061$-$7236375  & SMC/...    & 13.05 & 1  &  134  & 138.5 & ... & 3989 &    0.37 & 2.80 & $-$0.95 & $-$0.56 & $+$0.63 & $+$0.01 & $-$0.05 & $-$0.35 & $+$0.17 & $-$0.26 & $+$0.01 & $+$0.10 & $-$0.81 & $-$0.15 & $+$0.43 & $+$0.66 & $+$0.76 \\
2M00580325$-$7135564  & SMC/...    & 13.53 & 1  &  105  & 125.4 & ... & 3835 &    0.39 & 2.94 & $-$0.73 & $-$0.30 & $+$0.61 & $+$0.04 & $+$0.04 & $-$0.27 & $+$0.06 & $-$0.07 & $+$0.05 & $+$0.03 & $-$0.72 & $-$0.04 & $+$0.35 & $+$0.75 & $+$0.38 \\
2M00523564$-$7251053  & SMC/...    & 12.67 & 1  &  169  & 129.8 & ... & 4184 &    0.45 & 2.89 & $-$0.82 & $-$0.43 & $+$0.71 & $-$0.02 & $-$0.05 & $-$0.34 & $+$0.07 & $-$0.17 & $+$0.05 & $+$0.15 & $-$0.78 & $-$0.18 & $+$0.37 &    ...  &    ...  \\
2M01093482$-$7329422  & SMC/...    & 13.69 & 2  &  150  & 177.1 & 0.2 & 4009 &    0.51 & 2.95 & $-$0.85 & $-$0.30 & $+$0.56 & $+$0.06 & $+$0.10 & $-$0.26 & $+$0.15 & $-$0.11 & $+$0.02 & $+$0.01 & $-$0.82 & $-$0.04 &    ...  & $+$0.35 & $+$0.84 \\
2M00573246$-$7206024  & SMC/...    & 13.44 & 1  &  144  & 158.3 & ... & 4133 &    0.75 & 2.97 & $-$0.83 & $-$0.28 & $+$0.64 & $+$0.06 & $+$0.04 & $-$0.41 & $+$0.18 & $-$0.27 & $+$0.07 & $-$0.12 & $-$0.71 & $-$0.20 &    ...  & $+$1.58 & $+$0.96 \\
2M00540589$-$7208491  & SMC/...    & 13.63 & 1  &  149  & 116.8 & ... & 4047 &    0.83 & 2.88 & $-$0.79 & $-$0.41 & $+$0.66 & $+$0.05 & $-$0.02 & $-$0.36 & $+$0.12 & $-$0.20 & $-$0.00 & $+$0.13 & $-$0.71 & $-$0.06 & $+$0.19 & $+$0.72 & $+$0.90 \\
	\enddata
	\label{Table1}
\end{deluxetable*}


\end{document}